\documentclass[useAMS,usenatbib]{mn2e}
\usepackage{amssymb}
\usepackage{latexsym}
\usepackage{graphicx}

\def\f#1   {Fig.~\ref{#1}}
\def\s#1   {Sec.~\ref{#1}}
\def\tab#1   {Tab.~\ref{#1}}
\def\t#1   {Tab.~\ref{#1}}
\def\lum   {$\mathrm{L}_\mathrm{1.4GHz}$}
\def\comm#1   {{\tt (COMMENT: #1) }}

\def\sqdeg            {$\Box^{\circ}$}

\def\msol              {$\mathrm{M}_{\odot}$}

\def\wh                {W~Hz$^{-1}$}

\def\smo               {Smol\v{c}i\'{c}}

\title[HOD of radio AGN]{On the occupation of X-ray selected galaxy groups by radio AGN
  since $z=1.3^a$}

\author[V.~Smol\v{c}i\'{c} et al.]{V.~Smol\v{c}i\'{c}$^{1,2,3}$,
  A.~Finoguenov$^{4,5}$, G.~Zamorani$^{6}$, E.~Schinnerer$^{7}$,
 \\ \\ {\LARGE $\mathrm{M.~Tanaka^{8,2},\,  S.~Giodini^{9}, \, N.~Scoville^{10}}$}
        \\ \\
$^{a}$Based on observations with the National
Radio Astronomy Observatory which is a facility of the National Science
Foundation operated \\ under cooperative agreement by Associated Universities,
Inc.; the XMM-Newton, an ESA\\
  science mission with instruments and contributions directly funded by ESA
  Member States and NASA \\
$^{1}$ESO ALMA COFUND Fellow\\
$^{2}$European Southern Observatory, Karl-Schwarzschild-Strasse 2, 
D-85748 Garching b. M\"unchen, Germany\\
$^{3}$Argelander Institut for Astronomy, Auf dem H\"{u}gel 71, Bonn, D-53121, Germany\\
$^{4}$Max-Planck-Institut f\"ur Extraterrestrische Physik,
             Giessenbachstra\ss e, 85748 Garching, Germany\\
$^{5}$University of Maryland, Baltimore County, 1000
  Hilltop Circle,  Baltimore, MD 21250, USA\\
$^{6}$INAF - Osservatorio Astronomico di Bologna, via Ranzani 1, I-40127 Bologna, Italy\\
$^{7}$Max-Planck-Institut f\"{u}r Astronomie, K\"{o}nigstuhl 17, D-69117 Heidelberg, Germany\\
$^{8}$Institute for the Physics and Mathematics of the Universe, The University of Tokyo, 5-1-5 Kashiwanoha, Kashiwa-shi, Chiba 277-8583, Japan\\
$^{9}$Leiden Observatory, Leiden University, PO Box 9513, 2300 RA Leiden, the Netherlands\\
$^{10}$California Institute of Technology, MC 105-24, 1200
  East California Boulevard, Pasadena, CA 91125 
}

\begin{document}

\maketitle

\begin{abstract}
  Previous clustering analysis of low-power radio AGN has indicated
  that they preferentially live in massive groups. The X-ray surveys
  of the COSMOS field have achieved a sensitivity at which these
  groups are directly detected out to $z=1.3$. Making use of Chandra-, XMM- and
  VLA-COSMOS surveys we identify radio AGN members ($10^{23.6}\lesssim
  \mathrm{L_{1.4GHz}/(W\,Hz^{-1})}\lesssim10^{25} $) of galaxy groups
  ($10^{13.2}\lesssim\mathrm{M_{200}/M_\odot}\lesssim10^{14.4}$;
  $0.1<z<1.3$) and study i) the radio AGN -- X-ray group occupation
  statistics as a function of group mass, and ii) the distribution of
  radio AGN within the groups.  We find that radio AGN are
  preferentially associated with galaxies close to the center ($<0.2\cdot
  r_{200}$). Compared to our control sample of group members matched
  in stellar mass and color to the radio AGN host galaxies, we find a
  significant enhancement of radio AGN activity associated with
  $10^{13.6}\lesssim\mathrm{M_{200}/M_\odot}\lesssim10^{14}$ halos.
  We present the first direct measurement of the halo occupation
  distribution (HOD) for radio AGN, based on the total mass function
  of galaxy groups hosting radio AGN. Our results suggest a possible
  deviation from the usually assumed power law HOD model.  We
    also find an overall increase of the fraction of radio AGN in
    galaxy groups ($<1\cdot r_{200}$), relative to that in all
    environments. 
\end{abstract}


\section {Introduction}
\label{sec:introduction}
The general interest in radio AGN has increased in the last years as
their energy outflows both, theoretically and observationally
substantiate feedback processes highly relevant for massive galaxy
formation, and galaxy cluster/group physics (e.g.\ Croton et al. 2006,
Bower et al.\ 2006, \smo\ et al.\ 2009; \smo\ 2009,
\smo\ \& Riechers 2011, Giodini et al.\ 2009, 2010).  
Studies of low radio power (predominantly low-excitation, i.e.\ with weak or absent emission lines in their optical spectra) AGN have
shown that the probability of finding a radio AGN is a strong function
of stellar mass of the host galaxy (e.g.\ Best et al.\ 2005, \smo\ et
al.\ 2009). At high stellar masses ($M_*\sim10^{12}$~\msol ) this
probability approaches unity for galaxies out to $z=1.3$ (\smo\ et
al.\ 2009).  As high stellar mass galaxies usually reside in
group/cluster environments (e.g.\ Mandelbaum et al.\ 2006; Leauthaud
et al.\ 2010), it is still not entirely clear whether
it is the host galaxy and/or environmental properties that determine
the nature of radio AGN.

Initially,  Ledlow \& Owen (1996) found no
differences between the bivariate radio-optical luminosity function for
radio AGN in the cluster and in the field, suggesting that cluster
environment does not play a major role in radio AGN triggering. 
However, using a sample of radio AGN drawn from the NVSS
(NRAO VLA Sky Survey; \citealt{condon98}) and comparing the radio luminosity functions in (ROSAT
selected X-ray) clusters and in the field Lin \& Mohr (2007) found a
factor of 6.8 higher probability of a galaxy being a radio AGN in the
clusters than in the field 
($z<0.0647$.).

A detailed radio AGN clustering analysis at $z\sim0.55$ has been
performed by Wake et al.\ (2008). Constructing cross(auto)-correlation functions
of $0.4<z<0.8$ 2SLAQ Luminous Red Galaxies
(LRGs), NVSS radio detected LRGs and a control sample matched in redshift, color and
optical luminosity to the radio AGN they found that radio
AGN are significantly more clustered than the control
population of radio-quiet galaxies. Assuming various models for the halo occupation distribution, they find that radio AGN typically occupy more
massive halos compared to the control sample galaxies, suggesting that the
probability of finding a radio AGN in a massive galaxy (at
$z\sim0.55$) is influenced by the halo mass and/or cluster
environment.
 
It is important to stress that both clustering and stacking of halos,
used in the above analyzes (Wake et al. 2008, Mandelbaum et al. 2009)
suffer from degeneracies towards the distribution of radio AGN in
halos, either based on uncertainties of halo occupation of massive
galaxies (van den Bosch et al.\ 2004) or degeneracies of HOD modelling
(e.g. Miyaji et al.\ 2011). Thus, optimally one would directly study  the HOD by identifying the radio AGN within the galaxy groups.  
%
This can be done by using a well defined statistically significant sample of
radio AGN and X-ray selected galaxy groups and clusters. In this Letter we make use of such a sample in
the COSMOS field in order to directly study the halo occupation of
radio AGN.  We adopt $H_0=71$, $\Omega_M=0.25$, $\Omega_\Lambda=0.75$.

\vspace{-5mm}
\section{The samples}
\label{sec:data} 

The Cosmic Evolution Survey (COSMOS) is a panchromatic photometric and
  spectroscopic survey of 2\sqdeg\ of the sky observed by
  the most advanced astronomical facilities (e.g.\ VLA, Chandra, XMM-Newton, Subaru,
  CFHT; see Scoville et al. 2007). It allows a robust
source identification and extended selection of various source
types. The statistics on both radio AGN and selection of halos based
on their X-ray emission enables a construction of a statistically
significant catalog of radio AGN in massive halos, as described below.

\subsection{Radio AGN and red galaxy control samples}
\label{sec:radiodata}

We make use of the COSMOS radio AGN sample defined by \citet[see also
\citealt{smo09a}]{smo08}. It consists of $\sim600$ low-power
(\lum~$\lesssim10^{25}$~\wh ) radio AGN galaxies out to
a redshift of $z=1.3$. The sample was generated by matching the 20~cm
VLA-COSMOS Large Project sources (Schinnerer et al.\ 2007) with the
UV-NIR COSMOS photometric catalog (Capak et al.\ 2007). The selection required
optical counterparts with $i_\mathrm{AB}\leq26$, accurate photometry,
and a $\mathrm{S/N}\geq5$ (i.e.\ $\gtrsim50~\mathrm{\mu}$Jy) detection
at 20~cm. The radio AGN hosts were identified as red galaxies 
based on a rest-frame optical color 
($P1$ color $\geq0.15$) that mimics the
standard spectroscopic classification methods (see \citealt{smo06,smo08} for details).  The rest-frame color method efficiently
selects mostly type 2 AGN (such as LINERs and Seyferts), and
absorption-line AGN (with no emission lines in the optical spectrum),
while type 1 AGN (i.e.\ quasars, $\lesssim20\%$ of the total AGN
sample) are not included (see \citealt{smo08} for
detailed definitions). 

Furthermore, from the full COSMOS photometric redshift catalog
  \citep{ilbert09}, using the same optical magnitude, redshift and
  color selection criteria as for their radio-AGN, \citet{smo09a}
  generated a control galaxy sample from which the radio AGN are drawn
  ("control sample'' hereafter).
Here we make use of both of these samples, and add two additional
requirements. First, we use a luminosity limited sample of radio AGN
($23.6\leq\log{L_\mathrm{1.4GHz}/\mathrm{(W\,Hz^{-1})}}\lesssim25$;
$\mathrm{M_i}\leq -22.5$) in order to avoid selection effects induced
by limiting in flux rather than luminosity. Second, in order to
disentangle the dependence of radio emission on the host stellar mass,
we select a subsample of control galaxies ($\mathrm{M_i}\leq -22.5$)
that i) are not detected at 20~cm, and ii) have a stellar mass
distribution equivalent to that of our radio AGN. Given these two
  criteria we consider this sample to be
  ``radio-quiet'' and refer to it hereafter as "stellar mass matched
  control sample''.  Our criteria yield $217$ galaxies in the radio
AGN sample, $\sim5300$ in the control sample and $618$ in the stellar
mass matched control galaxy sample. The median redshifts of the radio
and stellar mass matched control galaxy samples are 0.72 and 0.85,
respectively.  At stellar masses exceeding $10^{12}~\mathrm{M_\odot}$,
we do not find galaxies without a radio AGN.

\subsection{X-ray sample of galaxy groups}
\label{sec:xdata}

The COSMOS field has been subject of intensive X-ray observations by
XMM-Newton (Hasinger et al. 2007; Cappelluti et al.\ 2009) and Chandra
(Elvis et al. 2009).  Following the method of Finoguenov et al. (2009)
a detailed subtraction of point sources was performed, which allowed
us to substantially reduce the contamination level of the catalog and
to improve the localization of the center of extended X-ray
emission. We use the complete set of XMM-Newton observations,
described in detail in Cappelluti et al.\ (2008). We also added the
data of the Chandra-COSMOS survey (Elvis et al.\ 2009), after the
removal of point sources. Identification of sources using the improved
COSMOS photometric redshift catalog \citep{ilbert09} is complete at
all redshifts. Furthermore, the zCOSMOS-BRIGHT (Lilly et al. 2009)
program provides spectroscopic identification for a dominant fraction
(67\%) of galaxies with $i_\mathrm{AB}\leq22.5$ and $z<1$. Most
  importantly, the mass-luminosity relation for the new sample of
  X-ray groups has been directly calibrated by Leauthaud et al.\
  (2010) using weak lensing. If the systematic uncertainties of groups
  hosting radio AGN are consistent with those of the entire sample,
  then the systematic uncertainties are within 10\% (Leauthaud et al.\
  2010).  The X-ray group catalog contains 210 systems at redshifts
below 1.3 with 130 groups having at least two spectroscopic redshifts
that match the redshift of the red sequence.

\subsection{Matching radio AGN and control samples to X-ray galaxy groups}
\label{sec:matching}

We have cross-correlated the X-ray galaxy group catalog  (see \s{sec:xdata} ) with the i)
full control sample, ii) stellar mass matched control sample, and iii) radio AGN sample (see \s{sec:radiodata} ). The correlation was done in
projected 2D space as accurate redshifts are assigned for all samples. We perform the matching for each radio AGN/control galaxy by searching
for the nearest cluster within a redshift slice centered at the galaxy's
redshift $z$. We take the half-width of the redshift slice to be
$\Delta z = 0.0334\cdot(1+z)$, which corresponds to $\sim3$ standard
deviations of the COSMOS photometric redshift distribution for $i<24$ (see Ilbert et al.\ 2009
for details). Prior to imposing a limit on the distance between the
radio AGN/control galaxy and the galaxy group center, below we consider the surface
number density of radio AGN inside groups.

\vspace{-5mm}
\section{Results}
\label{sec:results}

\subsection{Surface density profile of radio AGN in galaxy groups}
\label{sec:sd}

We derive the surface density profiles of the radio AGN and stellar
mass matched control samples satisfying our selection criteria
($\log{L_\mathrm{1.4GHz}}\geq23.6 \mathrm{[W/Hz]}$; $M_i\leq-22.5$;
$P1\geq0.15$). The profiles have been computed as distance to the
X-ray center in units of $r_{200}$\footnote{$r_{200}$ is defined as
  the radius within which the average density equals 200 times the critical
  density. Accordingly, $\mathrm{M_{200}}$ is the total mass embedded
    within $r_{200}$. The values for the COSMOS sample are taken from
    Finoguenov et al.\ (2009).} averaging over all the X-ray galaxy
groups in our sample (with photometrically masked regions taken into
account).  In \f{fig:sd} \ we show the surface density profiles for
our radio AGN and stellar mass matched control sample galaxies, scaled
by a factor of $2.85$ downward to account for the larger sample of the
second relative to the first.
There is a clear enhancement of occurrence of radio AGN at
group centers ($<0.2\cdot r_{200}$). 
This is consistent with results in the
local universe ($z<0.0647$; \citealt{lin07}), and will be further
discussed in \s{sec:discussion} .

\begin{figure}
\includegraphics[bb = 54 380 486 762, width=\columnwidth]{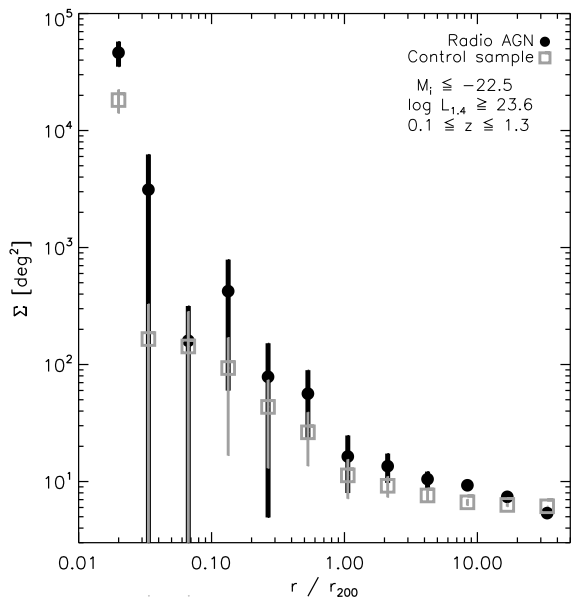}
    \caption{ The surface density distribution for our radio AGN
      (\lum~$\gtrsim4\times10^{23}$~W/Hz; filled dots), and stellar
      mass matched control
      (open squares) galaxies. Both samples are volume limited
      in the optical ($M_i\leq-22.5$). The profile for the stellar
      mass matched control
      sample has been  
      normalized by dividing it by a factor of 2.85 (corresponding to
      the ratio between the numbers of galaxies in the stellar mass
      matched control sample and the radio AGN sample).} \vspace{-4mm}
      \label{fig:sd}
\end{figure}


\subsection{Stellar mass function of radio AGN in galaxy groups: Effect of group environment on radio AGN triggering}

In the upper panel of \f{fig:mfstar} \ we show the stellar mass function computed using
the $1/\mathrm{V_{max}}$ method (see \smo\ et al.\ 2009 for details)
of both radio AGN and control galaxies (not matched in stellar mass) that occupy all environments
and galaxy groups ($\leq1\cdot \mathrm{r_{200}}$). It is noteworthy that in
the highest mass bin ($M_*\sim 10^{12}$~\msol ) essentially all red
galaxies are radio AGN that tend to occupy galaxy groups.  Dividing
the stellar mass function of radio AGN by that of the control galaxy sample yields the
volume corrected fraction of radio AGN relative to red host galaxies
in all environments and in groups. This is shown in the bottom panel of
\f{fig:mfstar} . These fractions can be considered as probabilities
that a massive red galaxy is a radio AGN in all and group
($\leq1\cdot \mathrm{r_{200}}$) environments, respectively. From
\f{fig:mfstar} \ it is obvious that the fraction of radio AGN is
enhanced in galaxy groups. In the highest mass bin ($10^{12.3}$~\msol ) the fraction of radio AGN is consistent with 1 irrespective of environment. At lower stellar masses ($10^{11.8}$ and $10^{12.3}$~\msol ) the enhancement of the fraction of radio AGN in groups is a factor of $\sim1.7$ and $1.6$ respectively.
 These results imply
that the triggering of radio AGN is directly linked to group environment.

\begin{figure}
\includegraphics[bb = 54 380 486 742, width=\columnwidth]{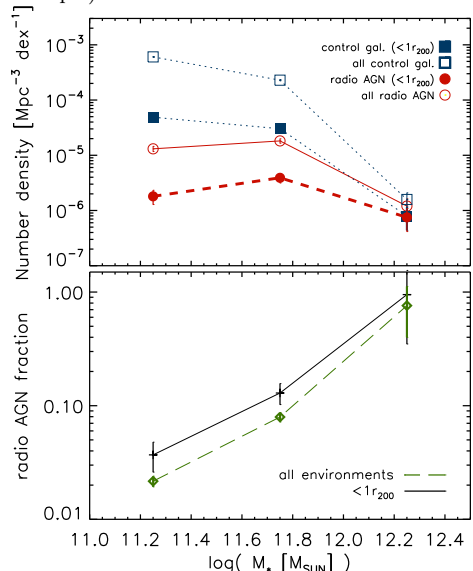}
\caption{
 Top panel: 
Top panel: stellar mass function for radio AGNs (red circles), red
control galaxies (blue squares) within $1\mathrm{r_{200}}$ of group centres (filled
symbols)
and in all environments (open symbols). Bottom panel: fraction of radio
AGNs (red circles) relative to red control galaxies in
groups ($\leq1\mathrm{r_{200}}$; solid line) and in all environments (dashed line).
  \label{fig:mfstar}}
\end{figure}

\begin{figure}
\center{ 
\includegraphics[bb = 84 415 450 642,
    width=\columnwidth]{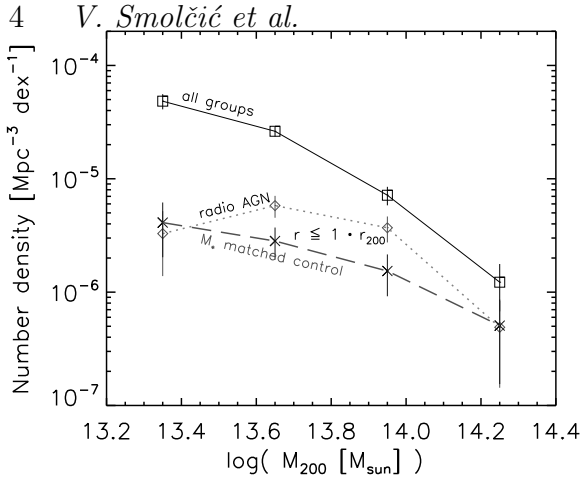}
\caption{
Differential mass function of X-ray galaxy groups. Squares
  indicate the full mass function. Diamonds show the mass
  function of groups hosting a radio AGN within
  $1\cdot r_{200}$. Crosses show the mass function of groups
  hosting a massive radio-quiet (stellar mass matched control sample) galaxy.
  \label{fig:mf}}
}
\end{figure}

\subsection{Radio AGN in galaxy groups: 
Halo mass function and halo occupation distribution}

In \f{fig:mf} \ we show the total mass function of all
galaxy groups, and those hosting a radio AGN or a stellar mass matched
control sample galaxy within $1\cdot r_{200}$.  Since the
radio sample is volume-limited, in calculating the mass function
we used the volume calculation ($1/\mathrm{V_{max}}$) for the X-ray
group sample. 

The shape of the total mass function of groups hosting a massive radio-quiet
(i.e.\ stellar mass matched control sample) galaxy within $1\cdot r_{200}$ differs from that
of all groups.  Massive radio-quiet galaxies tend to occupy higher
mass groups with increased frequency. 
On the other hand, groups hosting a radio AGN galaxy (\lum~$\geq10^{23.6}$~\wh ) show a very different distribution with a preference for high-mass halos ($\log(\mathrm{M_{tot}/M_\odot})\gtrsim13.5$), in particular in the range of $\log(\mathrm{M_{tot}/M_\odot})\sim13.5-14.1$, compared to the halo mass function of radio-quiet hosts.
%
This enhancement in number density (relative to a
stellar-mass matched sample) shows that the radio-AGN phenomenon (at
least in the given $\mathrm{M_{200}}$ range) must be governed by a
factor other than stellar mass. 

\begin{figure}
\center{ 
  \includegraphics[bb = 14 25 698 485,width=\columnwidth]{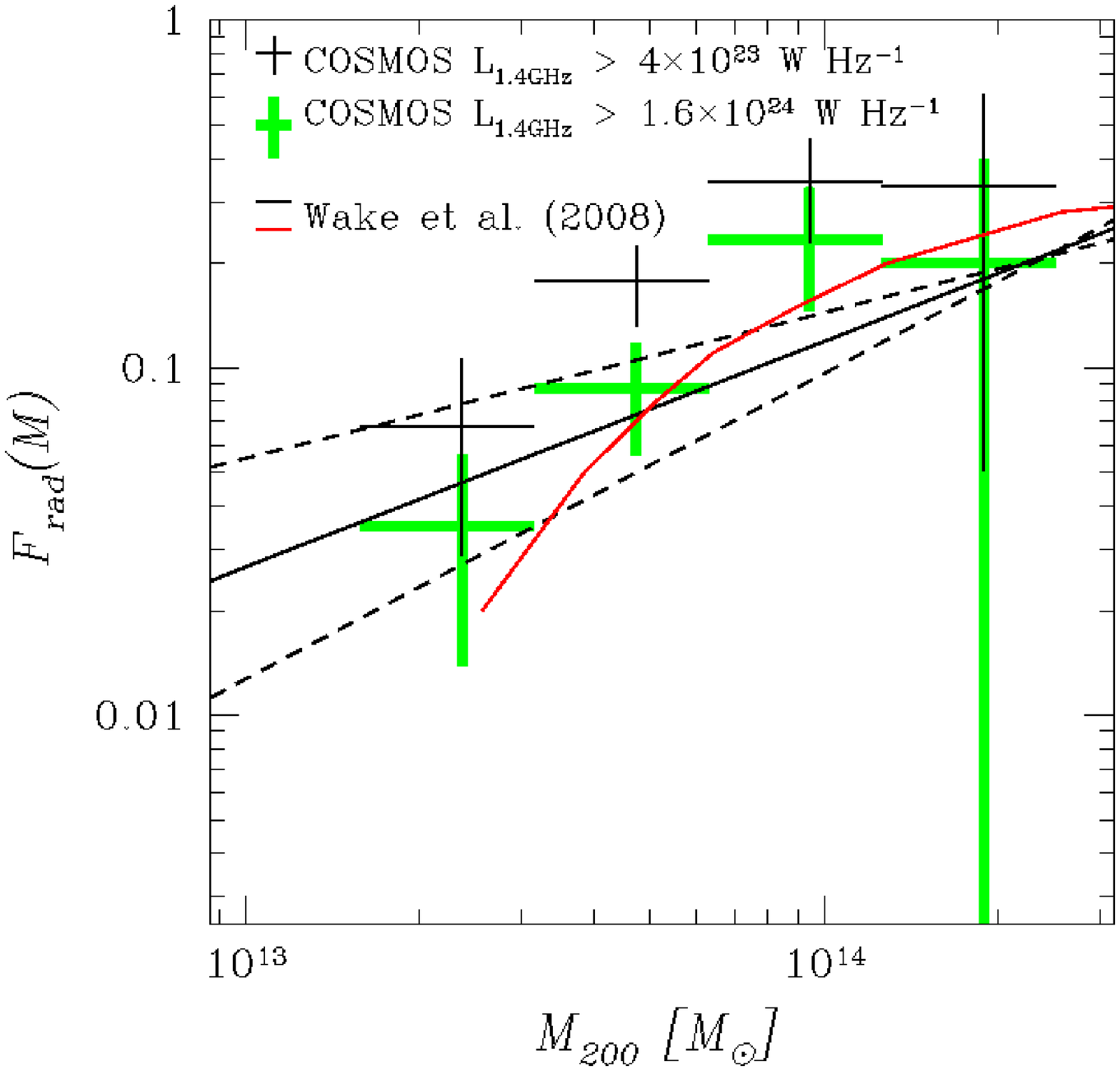}
  \caption{Occupation of X-ray selected groups by weak radio
    AGN. The COSMOS results (black: \lum~$\gtrsim4\times10^{23}$~\wh , gray: \lum~$\gtrsim1.6\times10^{24}$~\wh ) are shown as crosses, indicating the
    mass range on the X-axis and statistical (68\%
    confidence level) errors on the Y-axis.  The results of Wake et al. (2008), assuming a power-law dependence of radio-fraction, are shown as
    solid line (best fit) and short-dashed line (68\% confidence
    interval).  The solid curve shows the results of  Wake et
    al. (2008) obtained by dividing the best halo fits to the
    cross-correlation functions of their radio AGN and Luminous Red
    galaxies. \label{fig:hod}}
\vspace{-4mm}
}
\end{figure}

Our data-set allows us to generate the halo occupation distribution (HOD), which describes the occupancy of dark
matter halos by galaxies, 
directly. We simply need to divide the mass function of X-ray
groups hosting radio AGN within $1\cdot r_{200}$ by that of all
X-ray groups (see \f{fig:mf} ).  As shown in \f{fig:hod} , the HOD of
radio AGN ($\log{\mathrm{L_\mathrm{1.4GHz}}/\mathrm{(W\,Hz^{-1})}}\geq23.6$) within
$1\cdot r_{200}$ increases with total halo mass (embedded within
$r_\mathrm{200}$).  

It is conventional to fit the HOD with a power
law. The best fit to our data assuming an $\alpha ({M \over 10^{13.6}
  M_\odot})^\beta$ model yields $\alpha=0.16 (0.13-0.19)$ and
$\beta=0.90 (0.65-1.14)$, where the 68\% confidence interval is given in brackets.   The increase of the HOD
with halo mass implies that the occurrence of radio AGN in higher mass
halos is more frequent, compared to that in less massive ones. Such a
result is consistent with that found by \citet{wake08}.

\citet{wake08} have used a sample of 2SLAQ Luminous Red Galaxies
(LRGs) at redshifts 0.4-0.8, combined with FIRST and NVSS 20~cm radio
data to study the clustering properties of the i) full LRG sample, ii) radio
AGN with $\log{L_\mathrm{1.4GHz}}\geq24.2 \mathrm{[W/Hz]}$, and
iii) radio-quiet LRGs matched in color, redshift, and optical luminosity to
their radio-detected AGN. They fitted halo models to the constructed
auto- (full LRG sample) and cross-correlation (radio-detected and
matched LRGs) functions which allowed them to generate the radio AGN
HOD. In \f{fig:hod} \ we compare the HOD of \citet{wake08}, scaling their results to our mass definition and the value of the Hubble constant (Leauthaud et al. 2010).

The black (full and dashed) lines in \f{fig:hod} \ show their HOD modelling obtained from the best fit halo
model to their control sample clustering. In their fitting the radio AGN fraction was
left as a free parameter (assuming a power-law shape) to
match the clustering and space density of radio detected LRGs. The red curve also shown in \f{fig:hod} \ was constructed by Wake et al.\ in a different way, by simply dividing their halo fits to the radio AGN and control
sample data limited in luminosity.
%
The parameters of our HOD fit (assuming $\alpha ({M \over 10^{13.6}
  M_\odot})^\beta$, and using only  radio AGN more luminous
than $\sim1.6\times10^{24}$~\wh ) are $\alpha=0.09
(0.06-0.11)$ and $\beta=0.86 (0.45-1.22)$, in agreement with those derived by Wake et al.\ (2008).
The agreement in
the HODs suggests no significant redshift evolution between 0.55 and 0.73 (corresponding to the median redshifts of the samples analyzed in Wake et al.\ and here, respectively). Furthermore, the
deviant point at $\mathrm{M_{200}}\sim10^{14}$~\msol \ suggests a non-power law
form of the HOD as supported by the (red) curve in \f{fig:hod} \
(adopted from Wake et al.\ 2008).
With a lower limiting radio power ($\sim4\times10^{23}$~\wh ) we
find that the HOD overall increases by a factor of 2-2.5, with a
comparable slope to that given above. 



\vspace{-5mm}
\section{Summary and discussion}
\label{sec:discussion}

Using well defined samples of low-power radio AGN (which are mainly hosted by
massive early type galaxies; see e.g.\ \smo\ et al.\ 2008, 2009) and
X-ray groups in the COSMOS field at redshifts $z\leq1.3$, we analyzed
the occupation of galaxy groups ($10^{13.2}\lesssim\mathrm{M_{200}/M_\odot}\lesssim10^{14.4}$)
by radio AGN ($10^{23.6}\lesssim \mathrm{L_{1.4GHz}/(W\,Hz^{-1})}\lesssim10^{25}
$), as well as the effect of group environment on radio AGN. In order to investigate the latter, it is essential to
separate the stellar mass and clustering effects. The
blending of these two arises from the fact that i) the probability
of a massive early type galaxy to host a radio AGN strongly rises with
stellar mass (e.g.\ Best et al.\ 2005; \smo\ et al.\ 2009), and ii) massive red galaxies
preferentially reside in clustered environments (e.g.\ Mandelbaum et
al.\ 2006). To overcome this
bias, we have constructed a control sample of radio-quiet galaxies,
selected in the same way as our radio AGN, with a stellar mass
distribution matched to that of our radio AGN. Thus, any differences
in the derived distributions of these samples should arise due to
effects other than stellar mass. 

Generating surface density profiles for our radio AGN and radio-quiet
stellar-mass matched control galaxies as a function of $r_{200}$ (see
\f{fig:sd} ) we find a strong enhancement of radio AGN in group
centers ($<0.2\cdot r_{200}$). This is consistent with the results in
the local universe ($z<0.0647$; \citealt{lin07}).  
As our stellar mass matched control sample bypasses the
stellar mass/clustering bias (contrary to that in \citealt{lin07}),
our results directly link group environment to radio-AGN triggering.
The effect of groups on radio AGN is also clearly demonstrated in \f{fig:mfstar} , where we have shown that the fraction of radio AGN relative to red control galaxies is overall enhanced in galaxy groups.
%


Using the halo mass functions (derived from our X-ray data;
see \f{fig:mf} ) we generated the halo occupation distribution of
radio AGN (see \f{fig:hod} ). There is a very good agreement in the
HOD results between our survey and Wake et al.\ (2008;
\lum~$\gtrsim1.6\times10^{24}$~\wh ), despite very different
calculations. Generating auto/cross- correlation functions Wake et
al. (2008) estimated the mass of the halos based on the observed bias
(the discrepancy between the distribution of galaxies and the
underlying dark matter). In our analysis the mass of the halo has been
measured directly via weak lensing. Unlike the bias-mass relation,
weak lensing masses do not depend on the assumed growth of structure. 
Moreover, our results allow to break the degeneracy of HOD models and 
suggest a complex (non-power law) shape of the occupational distribution for radio AGN.

As our radio data are deeper than that used in \citet{wake08} we have
studied the HOD of radio AGN to lower radio powers
(\lum~$\gtrsim4\times10^{23}$~\wh \ which corresponds to a volume
limited sample out to $z=1.3$). With such a cut
the HOD of radio AGN overall increases by a factor of 2-2.5. The
  HOD can be interpreted as the probability that a massive red galaxy
  within $1\cdot r_{200}$ is a radio AGN. This probability can then be
  viewed as the fraction of time a massive galaxy spends as a radio
  AGN. Thus, it gives insights into the radio AGN duty cycle. Assuming
  that the red galaxy population was created at $z\sim2-3$ (i.e.\
  11~Gyrs ago) for \lum~$\gtrsim4\times10^{23}$~\wh\ this yields an
  average time a massive red galaxy in a galaxy group spends as a
  radio AGN of about 0.7~Gyr at $M_{200}\sim2\times10^{13}$~\msol\ to
  $\sim4$~Gyrs at the high mass end ($M_{200}\sim10^{14}$~\msol ). If
we restrain the elapsed cosmic time to the observed redshift range
($0<z<1.3$) we obtain an average time scale of $\sim0.6-3.5$~Gyrs. If
radio AGN triggering is caused by fueling of the supermassive black
hole via cooling of the large-scale hot gas \citet{churazov02}, then
the increase of the HOD with halo mass implies an overall higher
fueling efficiency in high mass halos.
In summary, our results yield that group environment enhances (on
average by a factor of $\sim2$) the probability of a red massive
galaxy being a radio AGN, and that radio AGN occupy higher mass halos
with increased frequency. 

\vspace{-7mm}
\section*{Acknowledgments}
\vspace{-2mm}
This researchwas was funded by the
European Union's Seventh Framework programme under grant agreement
229517; and contract PRIN-INAF
2007. The XMM--Newton project is supported by the Bundesministerium
fuer Wirtschaft und Technologie/Deutsches Zentrum fuer Luft- und
Raumfahrt (BMWI/DLR, FKZ 50 OX 0001), the Max-Planck Society.

\end{document}